\documentclass{llncs}
\newcommand{\e}{\varepsilon}

\usepackage{epsfig,amssymb,latexsym,color,amsmath,pifont}
\begin{document}

\pagestyle{empty}

\mainmatter

\title{Dynamical Regularization in Scalefree-trees \\ of Coupled 2D Chaotic Maps}

\titlerunning{Lecture Notes in Computer Science}

\author{Zoran Levnaji\'c}

\authorrunning{Z. Levnaji\'c}

\institute{Department for Theoretical Physics, Jo\v zef Stefan Institute \\
Jamova 39, SI-1000 Ljubljana, Slovenia\\
\email{zoran.levnajic@ijs.si} 
}

\maketitle

\begin{abstract}
The dynamics of coupled 2D chaotic maps with time-delay on a scalefree-tree is studied, with different types of the collective behaviors already been reported for various values of coupling strength \cite{ja4}. In this work we focus on the dynamics' time-evolution at the coupling strength of the stability threshold and examine the properties of the  \textit{regularization} process. The time-scales involved in the appearance of the \textit{regular state} and the \textit{periodic state} are determined. We find unexpected regularity in the the system's final steady state: all the period values turn out to be integer multiples of one among given numbers. Moreover, the period value distribution follows a power-law with a slope of -2.24.\\[0.2cm]
\textbf{Key words}: complex networks, coupled maps systems, emergent behaviour, self-organization in complex systems
\end{abstract}

\section{Introduction}

\textit{Complex networks} are overwhelmingly present in nature: a variety of systems from technology, biology or sociology can be seen as networks of interacting units that behave collectively \cite{dorogovtsev,boccaletti}. The paradigm of many basic elements that generate complex emergent behavior by interacting through the network links received a lot of attention as a convenient way to model complex systems. 

\textit{Coupled Maps Systems} (CMS) are networks of interacting dynamical systems (maps) that can be easily computationally modeled allowing the study of complex phenomena like synchronization, self-organization or phase transitions \cite{baroni,ahlers,zahera,bosarev}. The emergent behavior of a CMS can be investigated in relation to the architecture/topology of the network, type/strength of the coupling or the properties of the uncoupled units, often representing a model of applicational interest \cite{coutinho,rajesh}. The discovery of intrinsic modularity of many naturally occurring networks \cite{milo} triggered the investigations of the dynamical properties of small graph structures (termed \textit{motifs}) \cite{vega}, and their role in the global behavior of the network \cite{kahng,arenas}. Recent works emphasized the physical importance of \textit{time-delayed} coupling due to its ability to mimic the realistic network communication \cite{masoller,kurths}. Also, following the extensive studies of collective behavior in 1D CMS, the relevance of still poorly explored case of 2D CMS was recently indicated \cite{altmann,ja4}, having special relevance in the context of \textit{scalefree networks} \cite{dorogovtsev}.

In the previous works \cite{ja4,ja3} we studied the self-organization of coupled 2D standard maps on scalefree trees and motifs investigating different types of the collective behavior in function of the network coupling strength. After transients, the CMS dynamics was found to have two main collective effects: the \textit{dynamical localization}, acting at all non-zero coupling strengths that inhibits the chaotic phase space diffusion, followed by the \textit{regularization} acting at larger coupling strengths that produces groups of \textit{quasi-periodic} emergent orbits with common oscillation properties. For specific coupling values the network interaction eventually turns the quasi-periodic orbits into the \textit{periodic} orbits or even more complex structures like strange attractors, some of which unexpectedly appear to have characteristics of the \textit{strange non-chaotic attractors}, generally known to arise in the driven systems \cite{rama}. 

As we revealed, there exists a critical network coupling strength $\mu_c$ relatively common for all scalefree-trees and motifs, at which the collective motion becomes regular even after very short transient. In this work we will further examine the dynamics at $\mu_c$ for a fixed scalefree-tree. We will be concerned with the time-evolution of our CMS and the appearance of the collective effects in function of the initial conditions, seeking to reveal the relevant time-scales and the properties of the final emergent motion.

The paper is organized as follows: after defining our CMS on a scalefree-tree in the next Sect., we study its regularization process using the node-average and the time-average orbit approach in the Sect. \ref{The Average Orbit and The Dynamics Stabilization}. In the Sect. \ref{Properties of the Periodic Dynamical State} we study the appearance and the properties of the emergent periodic states, and conclude with the Sect. \ref{Conclusions} outlining the results and discussing the open questions.

\section{Coupled Map System Set-up on a Scalefree Tree} \label{Coupled Map System Set-up on a Scalefree Tree}
\begin{figure}[!hbt] \begin{center}
$\begin{array}{cc}
       \includegraphics[height=2.05in,width=2.4in]{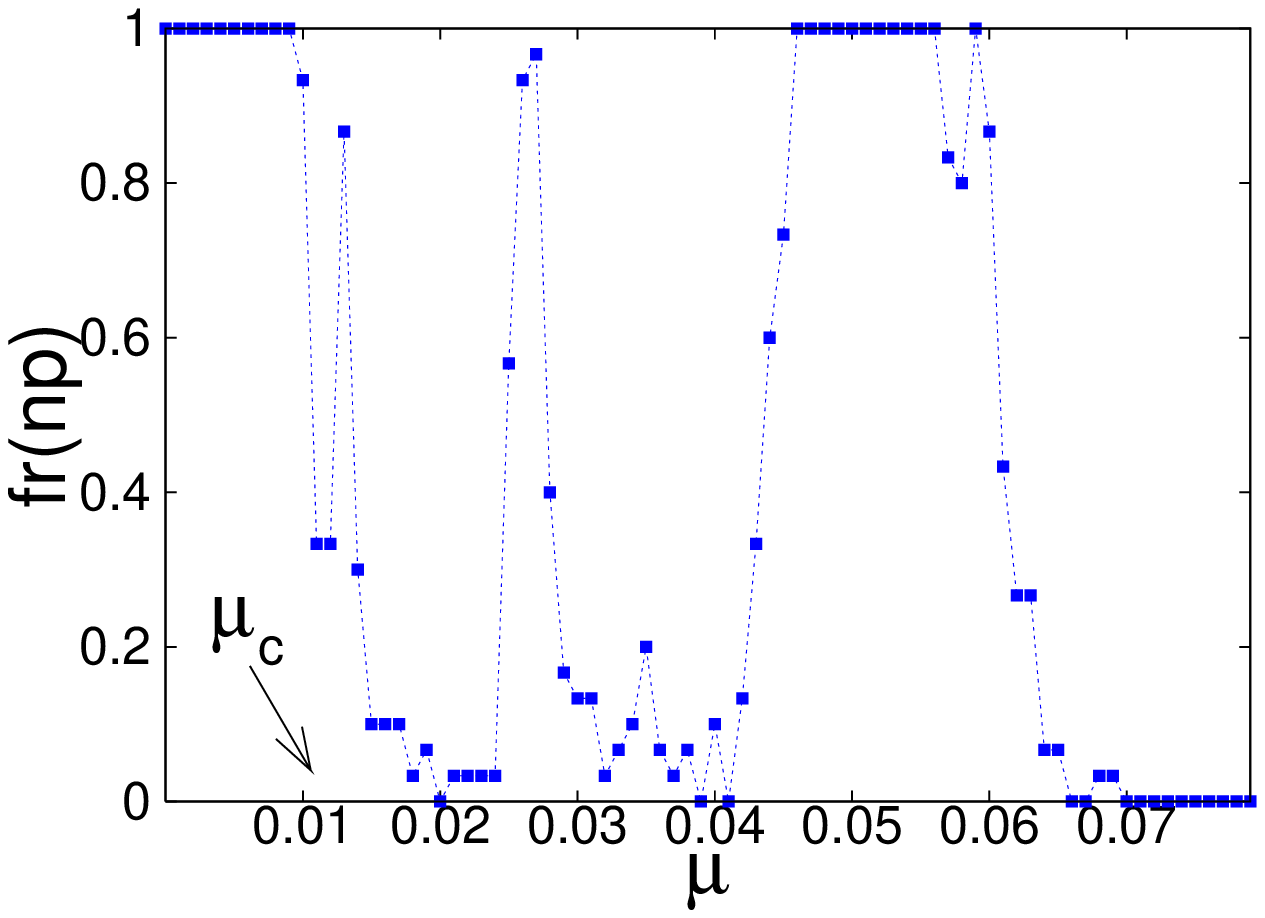} &
       \includegraphics[height=2.2in,width=2.25in]{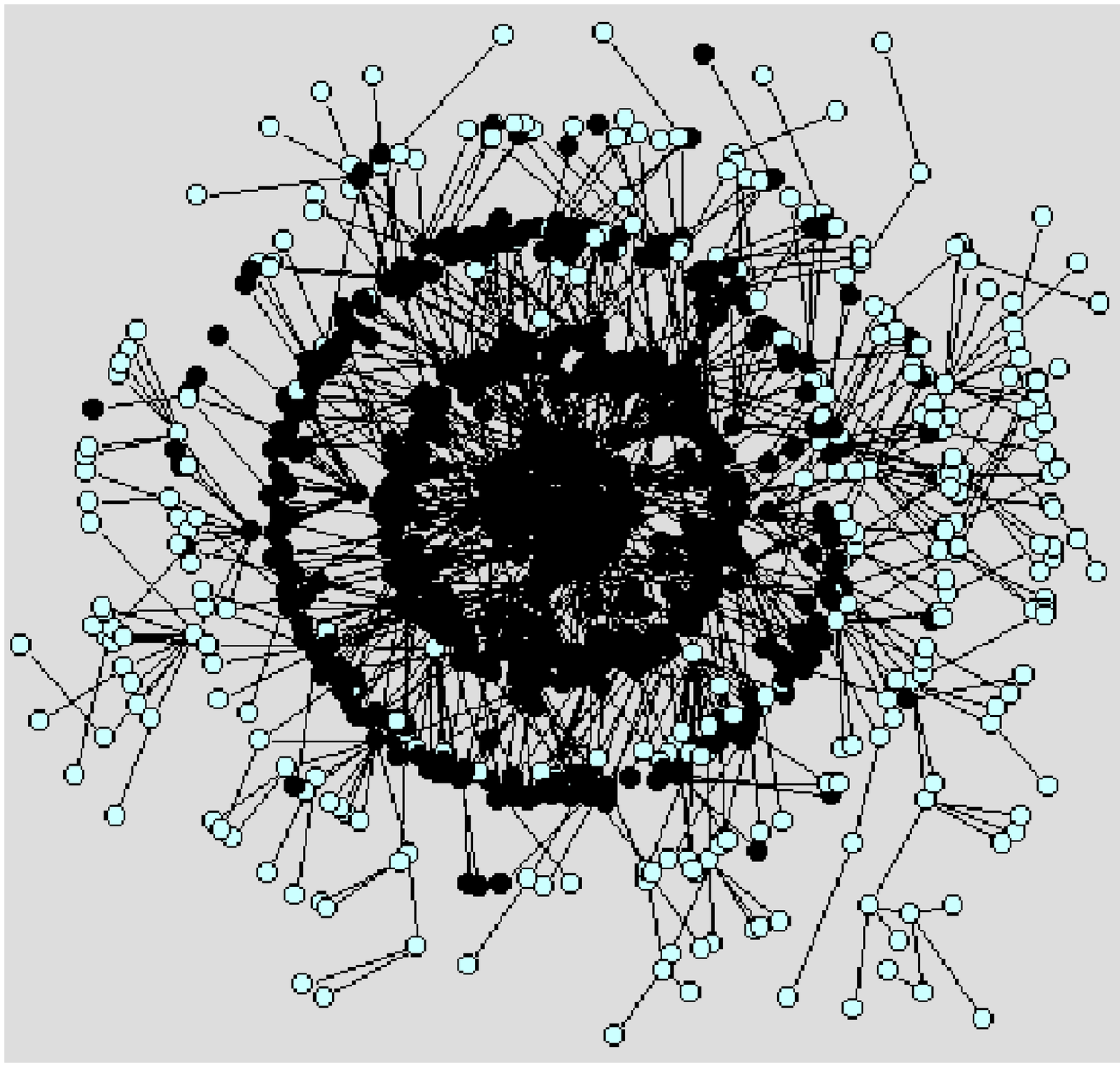} \\
       \mbox{(a)} & \mbox{(b)} 
\end{array}$ 
\caption{(a) Fraction of non-periodic orbits for an outer tree's node averaged over the initial conditions after a transient of $10^5$ iterations, (b) visualization of the tree studied in this paper. Bright nodes reached the periodic orbits after a transient of $5 \times 10^6$ iterations for a random set of initial conditions (see Sect. \ref{Properties of the Periodic Dynamical State}.).}
\label{intro}
\end{center} \end{figure}

A scalefree network with the tree topology is grown using the standard procedure of preferential attachment \cite{dorogovtsev} by 1 link/node for $N=1000$ nodes. Every node is assumed to be a dynamical system given by the Chirikov standard map: 
\begin{equation} \begin{array}{lllc}
  x' &=& x + y  + \e \sin (2 \pi x)  \;\;\;  &[mod \; 1]  \\
  y' &=& y + \e \sin (2 \pi x).   \;\;\; & 
\end{array} \label{sm} \end{equation}
The nodes (i.e. dynamical systems) are interacting through the network links by one-step time-delay difference in angle ($x$) coordinate so that a complete time-step of the node $[i]$ for our Coupled map system (CMS) is given by:
\begin{equation} \begin{array}{lll}
  x [i]_{t+1} &=& (1- \mu)x' [i]_t + \frac{\mu}{k_i} \sum_{j \in {\mathcal K_i}} 
(x[j]_t - x' [i]_t) \\
 y [i]_{t+1} &=& (1- \mu)y' [i]_t ,
\end{array} \label{main-equation} \end{equation} 
as in \cite{ja4,ja3}. Here, ($'$) denotes the next iterate of the (uncoupled) standard map (\ref{sm}), $t$ is the global discrete time and $[i]$ indexes the nodes ($k_i$ being a node's degree).  The update of each node is the sum of a contribution given by the update of the node itself (the $'$ part) plus a coupling contribution given by the sum of differences between the node's $x$-value and the $x$-values of neighboring nodes in the previous iteration (for motivation behind this coupling form, see \cite{ja4}). We set the standard map chaotic parameter to $\e=0.9$ and study the dynamics of CMS (\ref{main-equation}) on a fixed scalefree-tree (visualized in Fig.\,\ref{intro}b) for the fixed network coupling strength $\mu_c=0.012$. The $\mu_c$-value corresponds to the initial stability phase transition for this CMS with all the orbits becoming periodic after a sufficiently long transient, as shown in Fig.\,\ref{intro}a for a shorter transient (see \cite{ja4} for details). As opposed to the previous works, the focus will be maintained on the time-evolution of CMS (\ref{main-equation}) in relation to the initial conditions for a fixed $\mu_c$, rather then on the motion properties after transients for various $\mu$-values.

We therefore investigate the time-sequence of $2 \times N$-dimensional vectors 
\begin{equation} 
\left\lbrace  x[i]_t,y[i]_t \right\rbrace , \;\;\;\;\;  i=1,\cdots, N 
\label{orbit}\end{equation}
in function of the discrete time $t \ge 0$. The initial conditions are set by randomly selecting  the values $(x[i]_{t=0},y[i]_{t=0})$ for each node $[i]$ from $(x,y) \in [0,1] \times [-1,1]$. Future iterations are computed according to (\ref{main-equation}) with the parameter constraints described above. Updates of all the nodes are computed simultaneously for all the network nodes.
Besides considering the properties of a single node/single initial condition time-evolution sequence, we will also consider the quantities obtained by averaging over the tree's nodes, over the initial conditions or over the time-evolution, as they can provide alternative qualitative/quantitative insights into the time-evolution of the dynamics.

\section{The Average Orbit and The Dynamics Regularization} \label{The Average Orbit and The Dynamics Stabilization}

\begin{figure}[!hbt]
\begin{center}
$\begin{array}{ccc}
       \includegraphics[height=1.55in,width=1.56in]{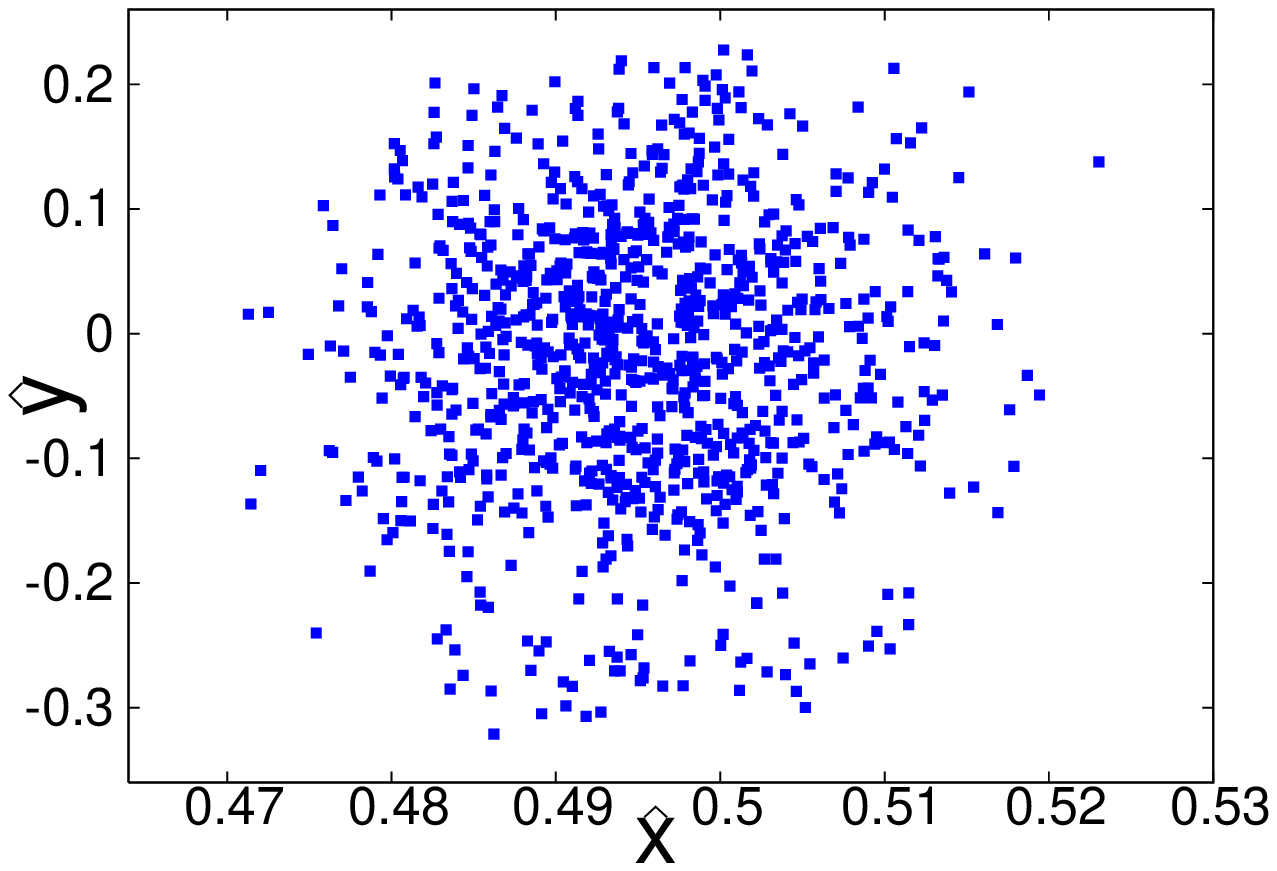} & 
       \includegraphics[height=1.55in,width=1.56in]{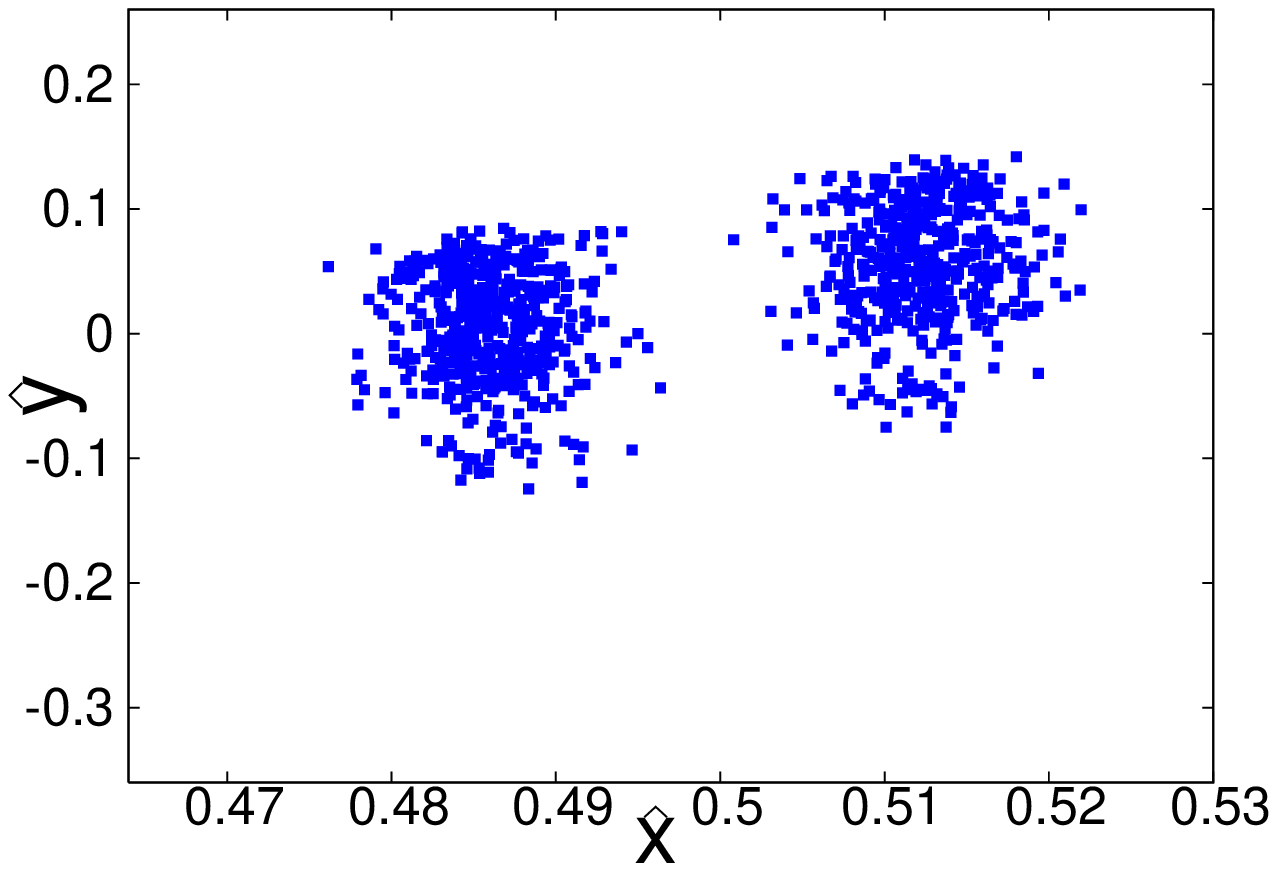} & 
       \includegraphics[height=1.55in,width=1.56in]{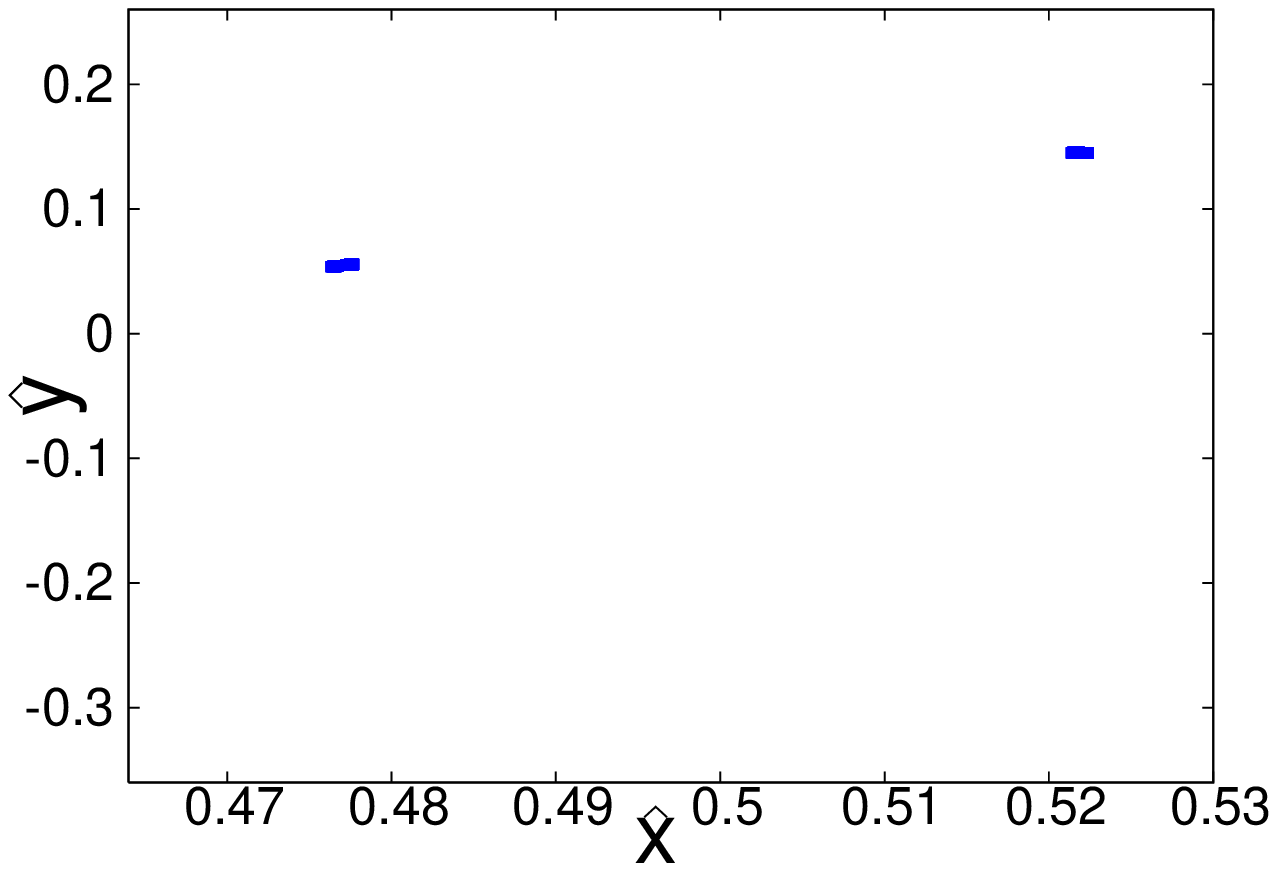} \\
       \mbox{(a)} & \mbox{(b)} & \mbox{(c)} 
\end{array}$ 
\caption{1000 iterations of a node-average orbit of the CMS (\ref{main-equation}) for a random set of initial conditions. (a) after 0, (b) after $5 \times 10^4$ and (c) after $5 \times 10^5$ iterations.}
\label{fig-tree-av-mu0012}
\end{center} 
\end{figure}

We begin the investigation of the general regularization properties of CMS (\ref{main-equation}) by examining the time-evolution of the \textit{node-average orbit} defined as:
\begin{equation} 
\left( \hat{x}_t,\hat{y}_t \right) = \frac{1}{N}\sum_{i=1}^{N} (x[i]_t,y[i]_t),
\label{naeo}\end{equation}
and representing the medium node position (averaged over the whole network) at each time-step. In Fig.\,\ref{fig-tree-av-mu0012} we show three stages of 
time-evolution of a generic node-average orbit of  (\ref{main-equation}). The initial chaotic cluster is divided into two smaller clusters that shrink and eventually become a quasi-periodic orbit (in the sense of not repeating itself while being spatially localized) oscillating between the clusters, with even and odd iterations belonging to the opposite clusters. The qualitative structure of this process is invariant under the change of the initial conditions. 

We will use the clusterization of this motion to design a more quantitative approach. Let us consider even and odd iterations of the node-average orbit $(\hat{x}_t,\hat{y}_t)_{even}$ and $(\hat{x}_t,\hat{y}_t)_{odd}$ as two separate sequences. They evolve in the phase space from initially being mixed together, towards each forming a separate cluster that shrinks in size. The sum of cluster sizes divided by their distance gives a good time-evolving measure of the motion's 'regularity'; one expects this quantity to decrease with the time-evolution, eventually stabilizing at some small value (or possibly zero). To that end we fix a certain number of iterations $\tau$ and coarse-grain the sequences by dividing them into intervals of length $\tau$ and considering two sequences of clusters (given as points within each interval for each sequence) instead. We compute the 2-dimensional standard deviations for each pair of clusters $\sigma_{even} (T)$ and $\sigma_{odd} (T)$, and measure the  distance $d_c (T)$ between their centers. These quantities depend on the coarse-grained discrete time $T$ given as the integer part of $(t/\tau)$. We define $\Sigma(T)$:
\begin{equation} 
\Sigma(T) = \dfrac{\sigma_{even} (T) + \sigma_{odd} (T)}{d_c (T)}, 
\label{sigma_approach}\end{equation}
that quantifies the evolution of the cluster separation process as a function of $T$. In Fig.\,\ref{fig-tree-clusterseparation}a we show the behavior of $\Sigma(T)$ for three different sets of initial conditions: 
\begin{figure}[!hbt]
\begin{center}
$\begin{array}{cc}
       \includegraphics[height=1.8in,width=2.15in]{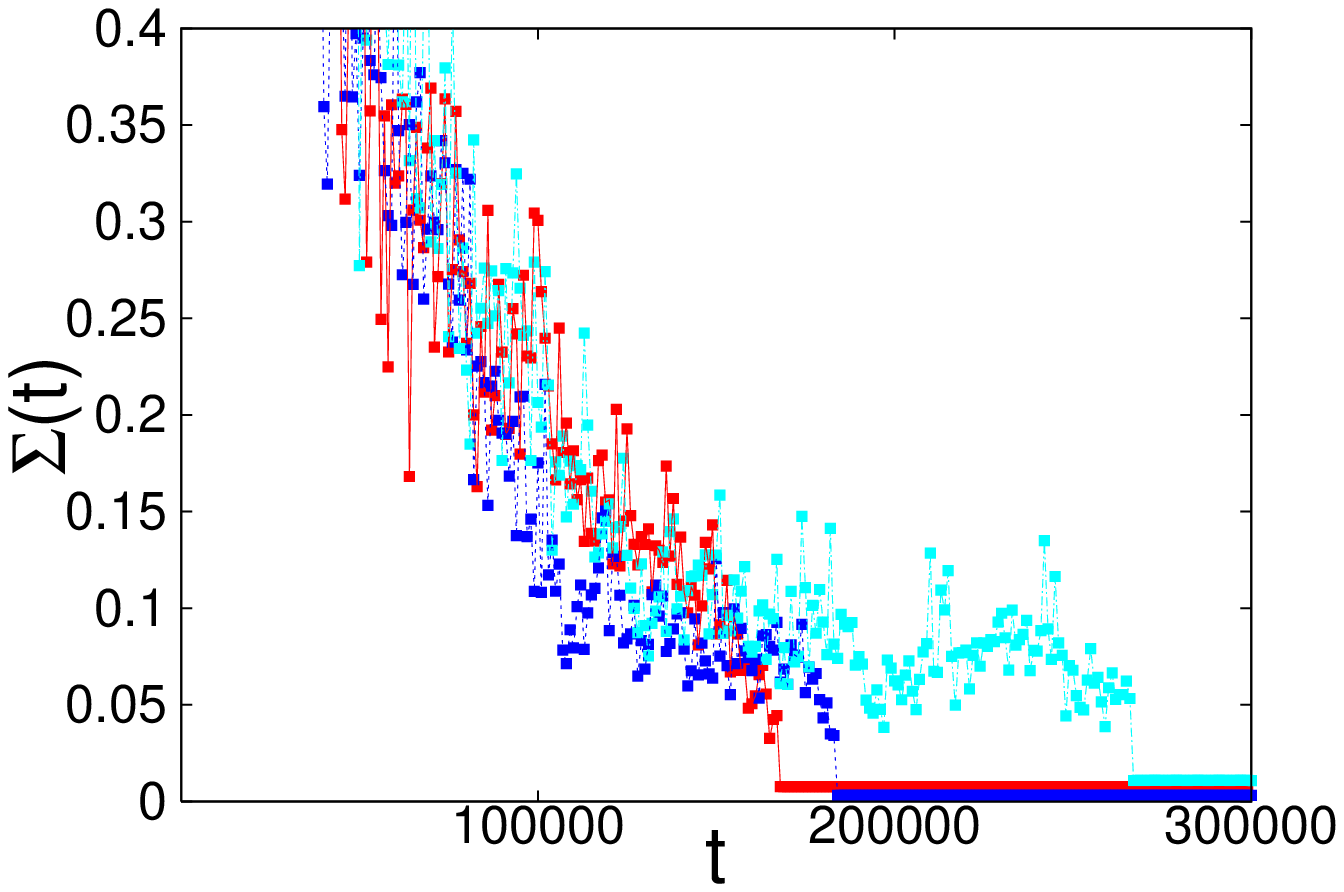} &
       \includegraphics[height=1.8in,width=2.15in]{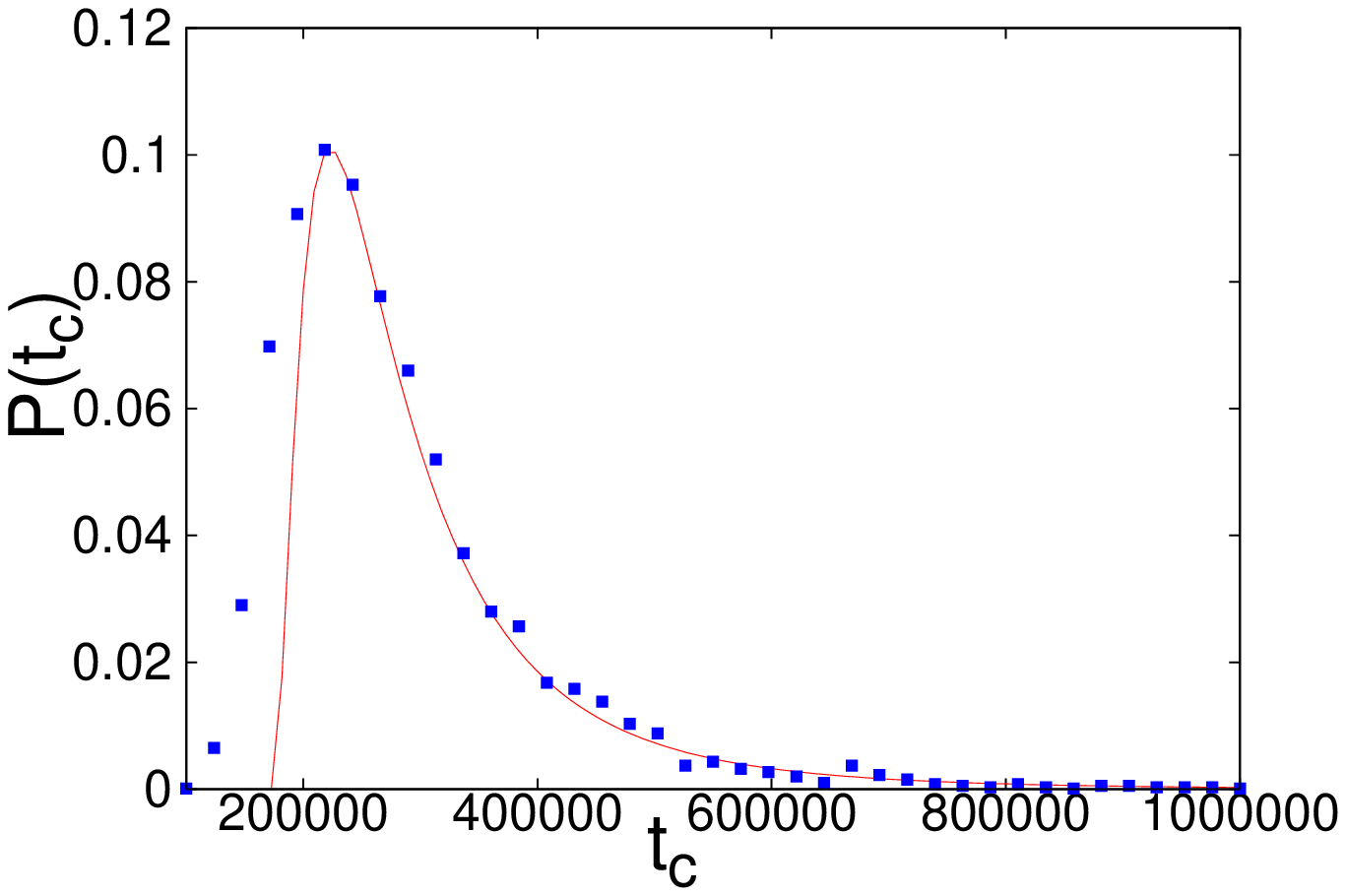} \\
       \mbox{(a)} & \mbox{(b)} 
\end{array}$ 
\caption{(a) Evolution of $\Sigma(T)$ for three sets of initial conditions (in normal time $t$), with time-window set to $\tau=1000$ iterations. (b) distribution of values of $t_c$ averaged over 1000 random initial conditions, with log-normal tail fit.}
\label{fig-tree-clusterseparation}
\end{center} 
\end{figure}
for each plot there exist a critical time $t_c$ at which the node-average dynamics suddenly 'regularizes' into a stable steady state characterized by a small $\Sigma(T)$-value that shows no further change with $T$. We term this collective dynamical state as \textit{regular} in the sense of constancy of $\Sigma(T)$ (Note: we use the term 'regular' only to make a clear distinction with the chaotic dynamics preceding this state, while a discussion of the precise properties of regular state will be given later). We emphasize that as this consideration is done using the node-average orbit, the result is a global property of scalefree tree CMS (\ref{main-equation}). The value of $t_c$ depends only on the initial conditions and it is given as the value of $T$ after which $\Sigma(T)$ remains constant.

In Fig.\,\ref{fig-tree-clusterseparation}b we examine the distribution of $t_c$-values over the initial conditions. The distribution $P(t_c)$ presents a log-normal tail with a prominent peak at  $\langle t_c \rangle \cong 2.79 \times 10^5$ iterations. Note that the value of $\langle t_c \rangle$ is by construction independent of the initial conditions and hence refers only to the network structure and the coupling strength $\mu_c$.

\subsection{Properties of The Regular Dynamical State} 

\begin{figure}[!hbt]
\begin{center}
$\begin{array}{cc}
\includegraphics[height=1.8in,width=1.9in]{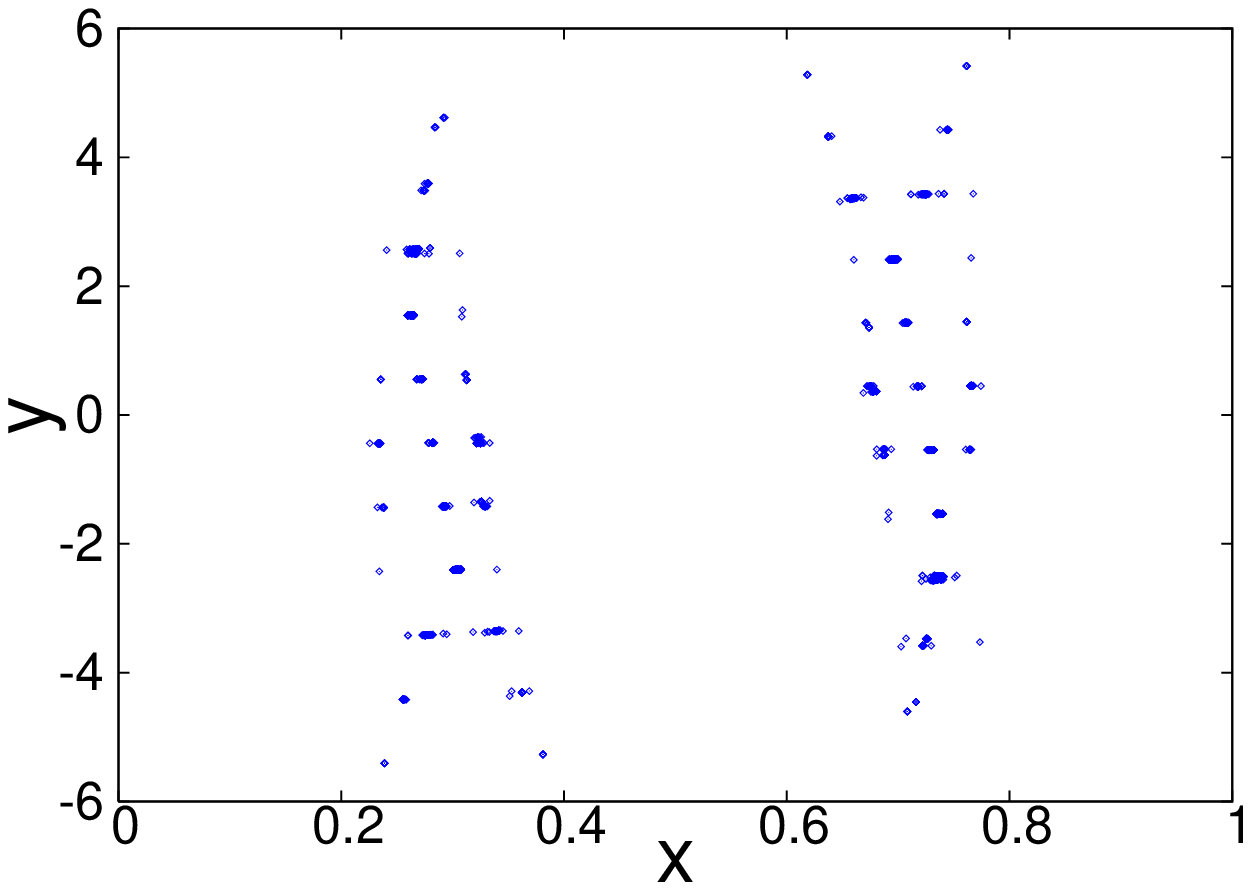} & 
\includegraphics[height=1.8in,width=2.4in]{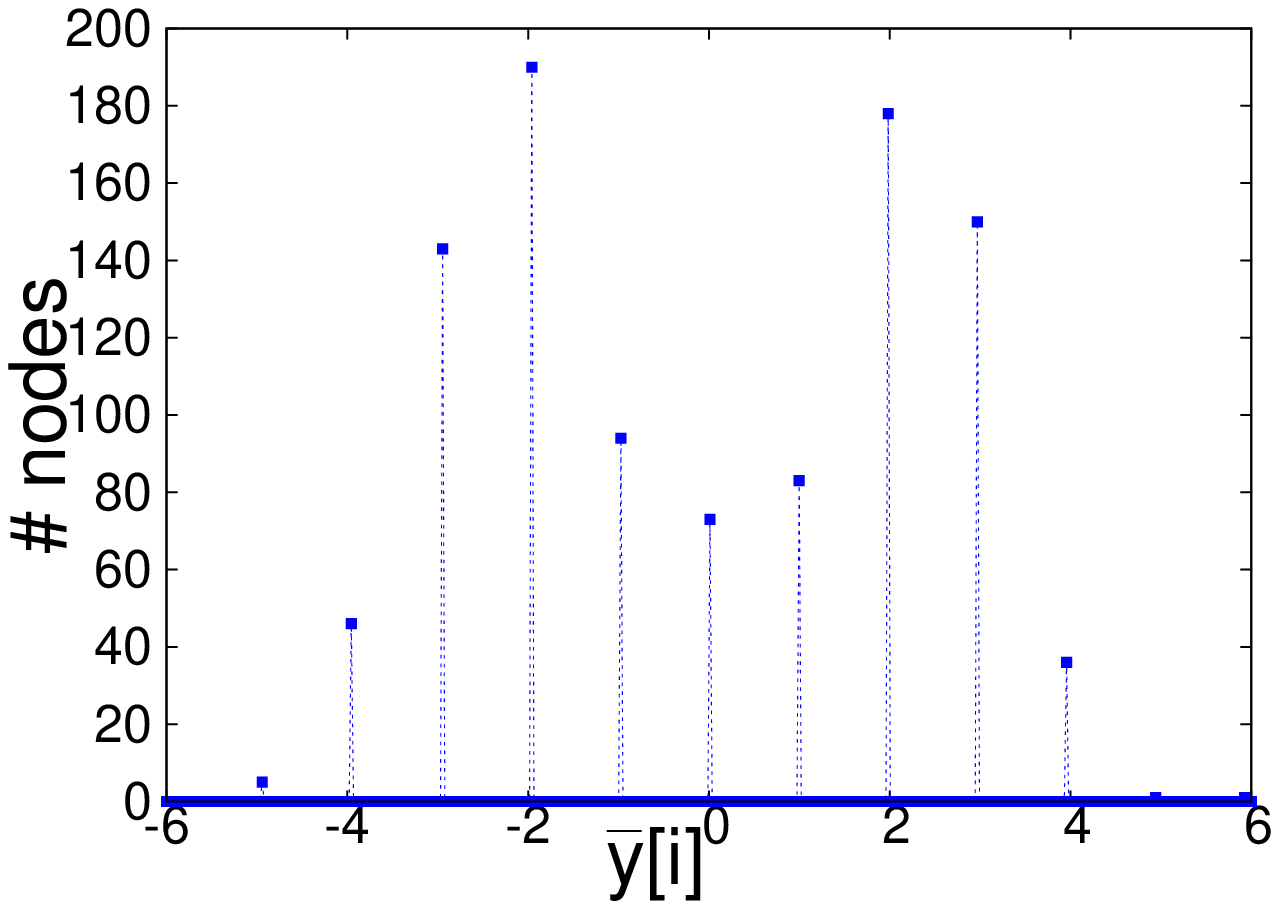} \\
\mbox{(a)} & \mbox{(b)} 
\end{array}$ 
\caption{A regular state of CMS (\ref{main-equation}) for a set of initial conditions after a transient of 500,000 iterations. (a) 10 iterations of all the nodes plotted in the same phase space, (b) distribution of $\bar{y}[i]$-values for all the nodes computed over $t_1-t_0=1000$ iterations.}
\label{fig-finalstate-tree-012-allnodes}
\end{center}
\end{figure}

The regular dynamics after $t_c$ is characterized by quasi-periodic oscillatory motion of each node  between two clusters, in analogy with the node-average orbit. All nodes however do not share the same clusters, but instead every node settles to oscillate between a certain pair of clusters that are horizontally corresponding to each other as illustrated in Fig.\,\ref{fig-finalstate-tree-012-allnodes}a. Every 
node therefore belongs to one group of nodes oscillating between a common pair of clusters with a phase space organization as in  Fig.\,\ref{fig-finalstate-tree-012-allnodes}a (see \cite{ja4}). Let the \textit{time-average orbit} of a node averaged over the time-interval $(t_0,t_1)$ be defined as:
\begin{equation} 
\left( \bar{x}[i],\bar{y}[i] \right) = \frac{1}{t_1-t_0}\sum_{t=t_O}^{t=t_1} (x[i]_t,y[i]_t),
\label{taeo}\end{equation}
which is a single phase space point that represents the node's average position during that time interval. In Fig.\,\ref{fig-finalstate-tree-012-allnodes}b we show the distribution of $\bar{y}[i]$-values for all the nodes of a single regular state: sharp peaks indicate groups of nodes sharing the same pairs of clusters and hence sharing the same $\bar{y}[i]$-value. Note that there are 11 pairs of horizontally linked clusters on Fig.\,\ref{fig-finalstate-tree-012-allnodes}a, in correspondence with the 11 peaks visible in  Fig.\,\ref{fig-finalstate-tree-012-allnodes}b. The symmetry properties of this distribution are invariant to the changes of the initial conditions.

\newpage
\section{Properties of the Periodic Dynamical State} \label{Properties of the Periodic Dynamical State}

Further evolution of any regular state eventually generates periodic orbits on all the tree's nodes under the dynamics of CMS (\ref{main-equation}). This is the final equilibrium steady state of this system that undergoes no further changes, and will be called the \textit{periodic state}. While the regular state is defined only qualitatively by the constancy of $\Sigma(T)$, the periodic state is precisely defined in terms of the periodicity of orbits on all the nodes. Group phase space organization of the nodes and their common oscillation properties described earlier ($\bar{y}[i]$-values, see Fig.\,\ref{fig-finalstate-tree-012-allnodes}) still remain, with the periodic orbits replacing the quasi-periodic ones.

\begin{figure}[!hbt]
\begin{center}
$\begin{array}{ccc}
\includegraphics[height=1.55in,width=1.63in]{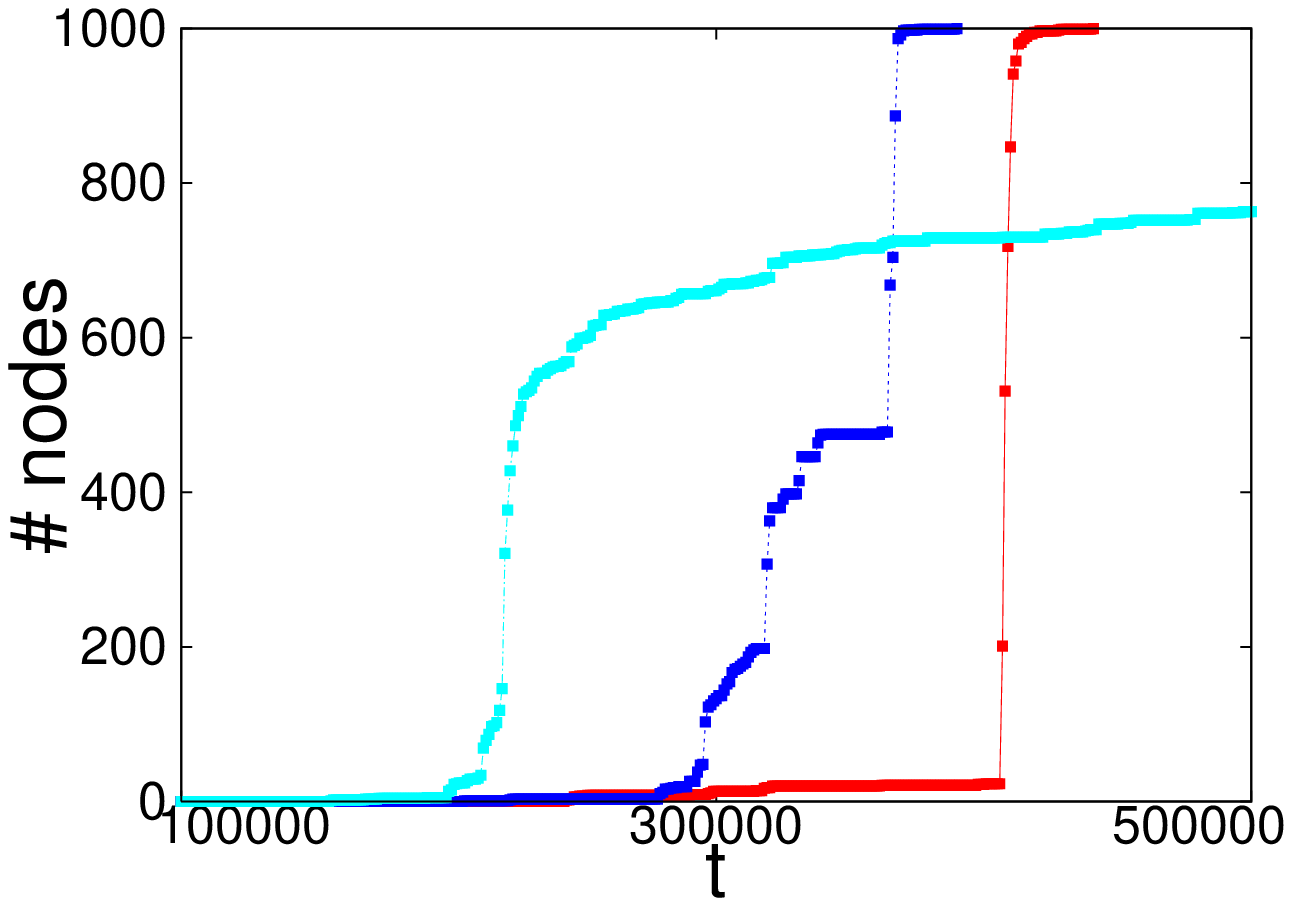} & 
\includegraphics[height=1.55in,width=1.51in]{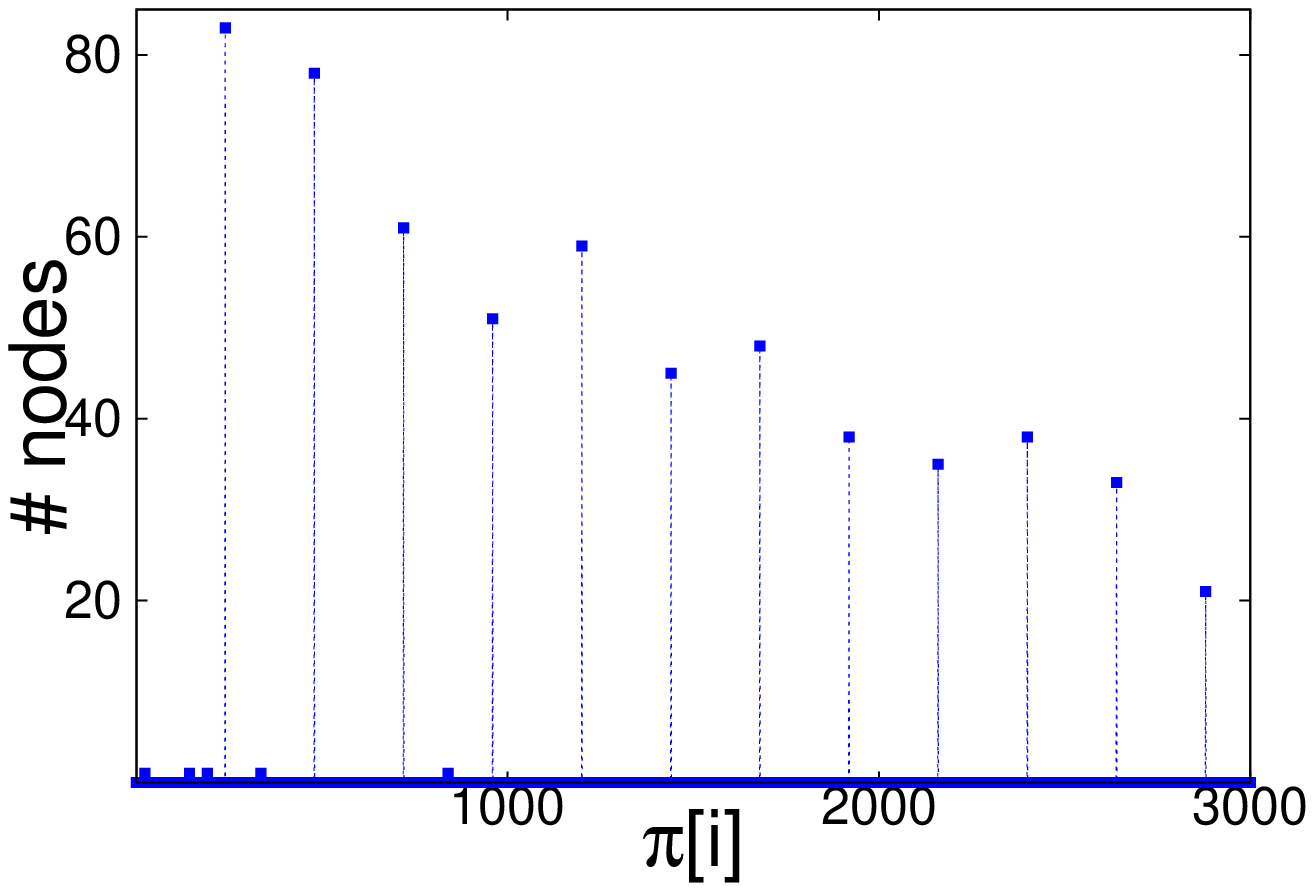} & 
\includegraphics[height=1.55in,width=1.55in]{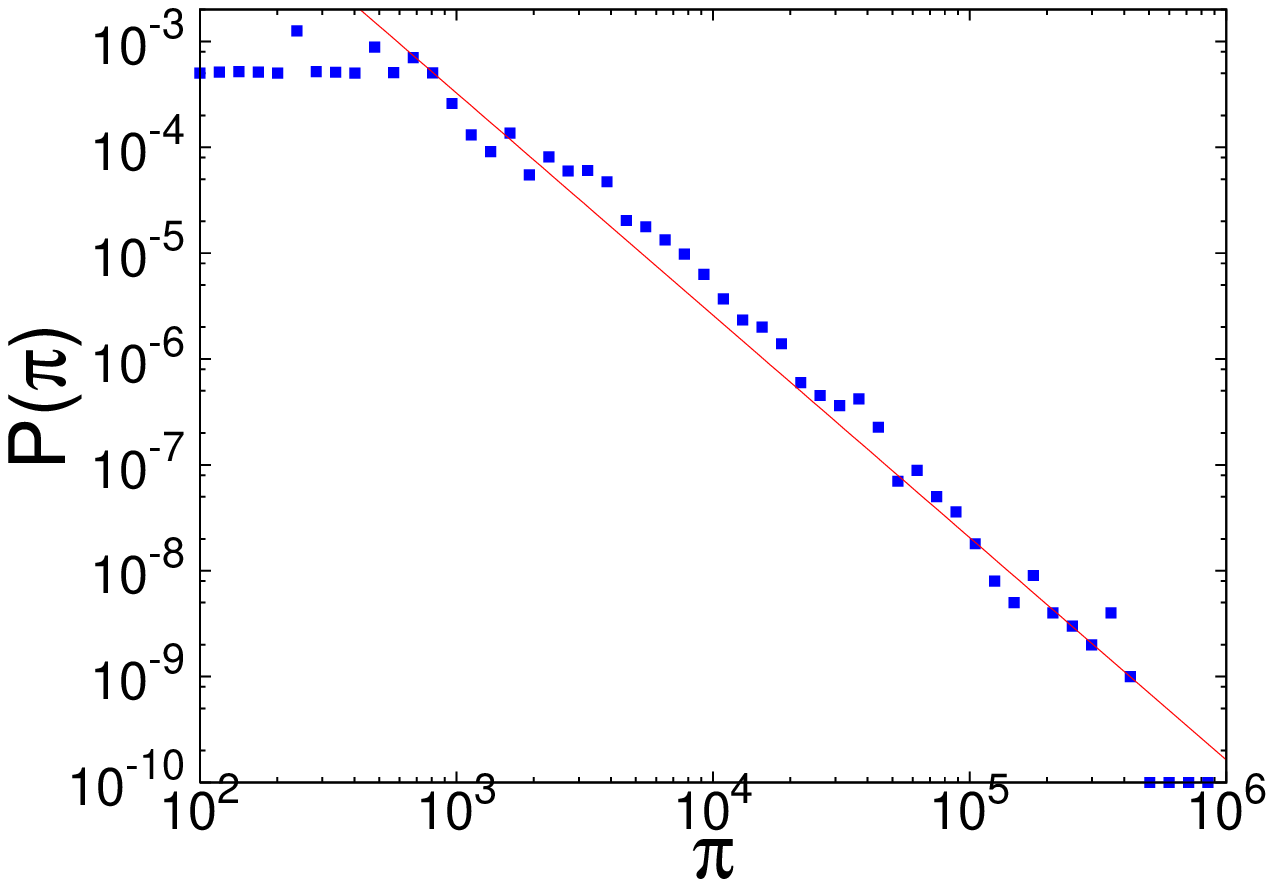} \\
\mbox{(a)} & \mbox{(b)} & \mbox{(c)} 
\end{array}$
\caption{(a) Number of nodes with periodic orbits as function of time for three sets of initial conditions, (b) histogram of period values (up to $\pi=3000$) for a single periodic state, (c) period distribution averaged over 20 periodic states, with slope of -2.24.}
\label{periodicstate}
\end{center}
\end{figure}

We describe the appearance of a periodic state in Fig.\,\ref{periodicstate}a: at each $\tau=1000$ iterations we check for nodes with the periodic orbits (for periods up to $\pi=10^5$) and report their number.
As opposed to the case of a regular state (Fig.\,\ref{fig-tree-clusterseparation}a), the nature of the transition to a periodic state is not unique. As clear from Fig.\,\ref{periodicstate}a besides a typical phase-transition (red), one can observe other phenomena including a slowly evolving non-equilibrium steady state (light blue) that does not reach an equilibrium state within $5 \times 10^5$ iterations (we analysed the states of this sort concluding they do not reach equilibrium even within $10^7$ iterations). Note that a sort of double phase-transition is also possible (dark blue) as the architecture of the tree may induce time scales into the regularization process. in Fig.\,\ref{intro}b shows a typical situation during the time-evolution: nodes achieve periodic orbits starting from ones with lesser links away from the hub, but not necessarily connected among them. Once the hub-node becomes periodic the whole tree is in a dynamical equilibrium, but the path of this transition can be diverse depending on the initial conditions.

\subsection{Properties of The Periodic Dynamical State} 

The key property of a periodic state is that almost all the period values turn out to be integer multiples of  a  given number. These \textit{base multiple numbers} are predominantly 240 (73\% of initial conditions) or 48 (18\% of initial conditions), with others being 96, 480, 720 etc. This is illustrated in Fig.\,\ref{periodicstate}b where we show a histogram of period values -- almost all the nodes have periods that are multiples of 240, with remaining nodes having periods multiples of 48. It is to be observed that all the base multiple numbers mentioned above are all multiples of 48. Furthermore, we computed the averaged distribution of period values (for 20 cases of multiples of 240) shown in Fig.\,\ref{periodicstate}c. The distribution exhibits a power-law tail with an exponent of about -2.24, for periods up to $\pi=10^6$.

These properties clearly indicate a presence of a self-organizational mechanism behind the creation of the periodic state of CMS (\ref{main-equation}). The same base multiple numbers are involved in cases of periodic states obtained for this CMS with other $\mu$-values, but with a different ratio of their presence depending on the initial conditions (generally, smaller the  $\mu$-value bigger the presence of larger base multiple numbers). Note that the periodicity of mentioned node orbits have no similarity with the periodic and quasi-periodic orbits known to exist in standard map's dynamics, as for this $\e$-value standard map shows only very little isolated regularity (otherwise being strongly chaotic).

\vspace*{-0.2cm}
\section{Conclusions} \label{Conclusions}

We examined the nature of the collective dynamics of CMS (\ref{main-equation}) with the fixed coupling parameters $\e=0.9$ and $\mu_c=0.012$ realized on a scalefree-tree, in terms of its time-evolution and the properties of its emerging dynamical states. We showed that for all the initial conditions dynamics becomes regular (in the sense defined above) after a critical time $t_c$. Also, for a majority of the initial conditions dynamics reaches a final steady state characterized by the periodicity of each node's orbit. Curiously, almost all the period values happen to be integer multiples of a given number that (depending on the initial conditions) varies within a given set of numbers. Also, for the periodic states having the base multiple number 240 (the most frequent one), the period value distribution follows a power-law with a tail slope of roughly -2.24.

Despite the oscillatory nature of emergent dynamics in many CMS known so far (mainly 1D), period values structure of this sort does not seem to be observed yet. We speculate that this self-organization feature might owe its origin to the time delayed structure of the coupling, and the nature of the standard map: once a node achieves a periodic orbit, its neighbours ought to do the same, inducing the correlation in the period values.

Open questions include network organization of periodic states in terms of period--node relationship, as well as investigation of the steady states of this CMS realized with different networks. It might be also interesting to study similar CMS with time-delays that vary throughout the network in a way to model a given naturally occurring behaviour. Of particular interest might also be the investigation of the non-equilibrium steady state mentioned in Fig.\,\ref{periodicstate}a (light blue) from a statistical point of view. \\[0.5cm]
{\bf Acknowledgments.} This work was supported by the Program P1-0044 of the Ministery of Higher Education, Science and Technology of Republic of Slovenia. Many thanks to prof. Bosiljka Tadi\'c for her guidance and useful comments.
\bibliography{levnajic-references}
\end{document}